\documentclass{article}
\usepackage[draft]{hyperref}
\usepackage{amsmath}
\usepackage{graphicx}
\usepackage{amsfonts}

\begin{document}

\title{All-optical Switch Based on Optical Waveguide Coupling with Micro Cavity Array}

\author{Bin Yang$^{1, 2, *}$, Heling Li$^{1}$, Hongshen Zhao$^{1}$, Tao Yang$^{1}$ \\$^1$School of Physics and Infermation Science in Ningxia University,Yinchuan 750021,China\\$^2$Department of Physics in Beijing Normal University, Beijing 100875, China\\$^*$Corresponding author: ybin92@gmail.com}

\maketitle

\begin{abstract}
This paper theoretically analyzes the optical transmission characteristics of an optical waveguide when coupling to a micro cavity array. The results showed that not only were there sharp peaks on the transmission and reflection spectra, but also that a certain system configuration can produce a backward wave to obtain a   phase shift in the small detuning range between the incident wave and the micro cavities. Based on a discussion of the relationship between the transmission and the number and the dissipation coefficients of each coupled micro cavity, the paper proposes a high efficiency all-optical switch model.
\end{abstract}

\section{Intoduction}
The all-optical switch is the one of the most basic elements in an optical network. Therefore, high efficiency optical switches \cite{englund2007controlling, bajcsy2009efficient, dawes2009optical, wen2011all, abaslou2012compact} provide a research hotspot, because of their applications in optical integration. Most current optical switch models are based on the principle of interference with the signal light after obtaining a $\pi$ phase shift through optical switching devices by controlling the optical signal, so that the switching of the signal light can then be controlled. This type of optical switch device needs to be long enough to obtain the   phase shift, while because of the limitations of the properties of optical materials, this is not helpful for device integration. Another type of optical switch \cite{popov2005nonlinear, myslivets2002nonlinear} made from ordinary nonlinear materials requires a certain optical power threshold, which cannot be applied in weak light environments. In recent years, optical switches using photonic crystals and nanometer-scale micro cavities have received wider attention \cite{bajcsy2009efficient, cuesta2004all, barclay2006integration}. These switches mainly use the slow light effect of the photonic crystal \cite{baba2008slow} or the resonance characteristics of the micro cavity, which not only maintains the small physical size of the devices, but also achieves the signal light switching function over a wider frequency band range.

The slow light effect of photonic crystals is similar to the resonance characteristics of the micro cavity \cite{krauss2007slow}. Therefore, this paper analyzes the optical transmission characteristics of the optical waveguide when coupling to the micro cavity array in detail using cavity quantum electrodynamics theory (cavity-QED) \cite{vahala2004optical}. The analysis results of the study also indicated that a certain number of micro cavity-coupled optical waveguide models can be used as a highly efficient optical switch. The analysis of our model can also provide a theoretical basis and method for optimization of this type of optical switch.

\section{Model and Theoretical Analysis}
  \begin{figure}[htbp]
   \centering
   \includegraphics[width=8.3cm]{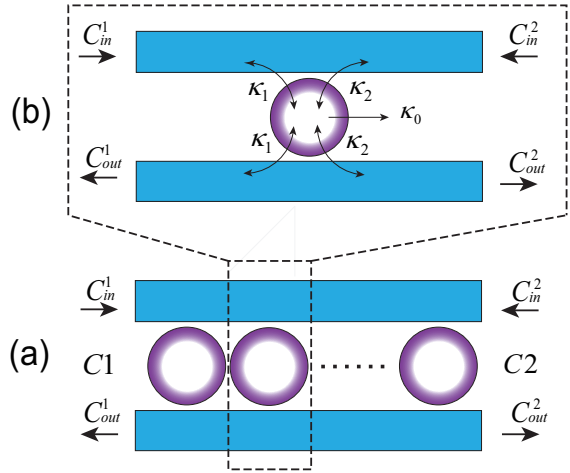}
        \caption{Model of optical waveguide coupling to micro cavity array. (a) The whole system with N micro cavities; (b) A single unit of the system. }
    \label{1}
  \end{figure}

The structure of the system model studied in the present paper shows that the $N$ micro cavities are coupled with the optical waveguide L (Fig. 1(a)). Figure 1(b) represents a single-unit structure as part of the entire system. Consider each micro cavity to be a single-mode cavity, with their natural frequencies represented by $\omega_0 $. The other parameters in the model are explained as follows: $k_0$ refers to the dissipation coefficient of the ith cavity coupled to the external environment, and its relationship to the quality factor of the micro cavity $Q_0$ is given by $k_0 = 2 \omega_0 Q^{-1}_0$, $k_1$ and $k_2$ are the dissipation coefficients in the left and right directions, respectively, resulting from the ith cavity coupled to the waveguide. Each micro cavity is coupled with a waveguide through   and  , rather than direct coupling between each micro cavity. $C^{1}_{in}$ and $C^2_{in}$ refer to the light wave flow operators of the ith unit input from the two sides of the waveguide, and  $C^1_{out}$  and  $C^2_{out}$  refer to the light wave flow operators of the ith unit output from the two sides of the waveguide.

First, analysis of the optical transmission characteristics of a single unit was conducted. Under weak excitation conditions, the Langevin motion equation of the annihilation operator  $a(t)$  in the time domain in the ith micro cavity is given by\cite{walls2008quantum}:  
\begin{equation}
\label{eq:crosseq-1}
  \frac{da(t)}{dt} =-i\omega_0a(t)-\frac{k_0+k_1+k_2}{2}a(t)-
            \sqrt{k_1}C_{in}^1-\sqrt{k_2}C_{in}^2
\end{equation}

which is a first order linear inhomogeneous differential equation based on $a(t)$ . The equation can be solved by using a Fourier transform, in which the annihilation operator within the frequency domain, $\tilde{a}(\omega)$  , is obtained and is given by:
\begin{equation}
\label{eq:crosseq-2}
   \tilde{a}(\omega)=\frac{\sqrt{k_1}\tilde{C}^{1}_{in}(\omega)+
       \sqrt{k_2}\tilde{C}^2_{in}(\omega)}{i \delta\omega-\gamma}
\end{equation} 

where $\gamma=(k_0+k_1+k_2)/2$, $\omega$ is the frequency of the incident wave, and  $\delta\omega= \omega -\omega_0$  is the detuning of the incident wave and the micro cavity. When applied to the input and output relationship for the cavity and waveguide coupling \cite{PhysRevLett.92.117902},$C^{i}_{out}=C^{i}_{in}+\sqrt{k_i}a(\omega)$ , where $(i=1,2)$,  $a(t)$  can be removed from Eq. (2) to obtain the input and output relationship for the ith unit (for writing convenience, the light wave flow operator $\tilde{C}(\omega)$  and is simply written as $C$  in the following statements):
\begin{equation}
\label{eq:crosseq-3}
  \begin{matrix}
  \begin{bmatrix} C^2_{in} \\ C^2_{out} \end{bmatrix}
  \end{matrix}
  = T_i
  \begin{matrix}
  \begin{bmatrix} C^1_{in} \\ C^1_{out} \end{bmatrix}
  \end{matrix}
\end{equation}

In the above equation,  $T_i$ refers to the transmission matrix of the ith unit. Its specific form is given by:
\begin{equation}
\label{eq:cross-4}
T_i = \lambda\begin{matrix}
     \begin{bmatrix}  k_1 & \gamma - i\delta \omega \\
          \gamma - 2\sqrt{k_1k_2} -i\delta\omega & -k_2 \\
     \end{bmatrix}
     \end{matrix}
\end{equation}
Where the coefficient
$\lambda=(\gamma-\sqrt{k_1k_2-i\delta \omega})^{-1} $.

The whole system, with the N micro cavities coupled to the optical waveguide, is now considered. For convenience, we assume that the optical path difference between two adjacent units is an integral multiple of the vacuum wavelength. Figure 1(a) shows that $C^2_{out}$ and $C^2_{in}$  of the ith unit are equal to  $C^1_{in}$ and $C^1_{out}$ of the i+1th unit. The total transmission relationship of the system can then be obtained by using Eq. (3):
\begin{equation}
\label{eq:crosseq-5}
  \begin{matrix}
  \begin{bmatrix} C^2_{in} \\ C^2_{out} \end{bmatrix}
  \end{matrix}
  = T_{t}
  \begin{matrix}
  \begin{bmatrix} C^1_{in} \\ C^1_{out} \end{bmatrix}
  \end{matrix}
\end{equation}

where $C_{in}^1$,  $C_{out}^1$, $C_{in}^2$, and $C_{out}^2$ are the input and output flow operators on the two sides of the system. The total transmission matrix of the system is  $ T_{t} = T_n T_0
T_{n-1} \ldots T_0 T_1 $ , where  $T_i$ ($i$ = 1,2, \ldots, n) is the transmission matrix of the ith micro cavity unit, shown as Eq. (4).  $T_0$ is given by
\begin{equation}
  T_0 = 
  \begin{matrix}
  \begin{bmatrix} 0 & 1 \\ 1 & 0 \end{bmatrix}
  \end{matrix}
\end{equation}

The reflection coefficient of the system is defined as: $r=\sqrt{R}exp(i\phi)=\frac{<C_{out}^2>}{<C_{in}^1>}$, where $R$ refers to the reflection ratio and  $\phi $ is the phase shift of the reflected wave relative to the incident wave. Similarly, the system transmittance is defined as£º$t=\sqrt{T}exp(i\phi)=\frac{<C_{out}^1>}{<C_{in}^1>}$ where $T$ refers to the transmittance and  $\phi$  is the reflected wave phase shift relative to the incident wave. These two definitions are used under the conditions where single or multiple micro cavities are coupled to the waveguide.

\section{Analysis of Transmission Characteristics}
First, we assume that only one micro cavity is coupled to the waveguide, and choose a dissipation coefficient that is normalized with respect to $k_1$ (unit: THz), where $k_2 = 0.001k_1 $, and $k_0=0.005k_1 $. The input light wave signal is set as $C^1_{in}$ only, i.e., $C^2_{in} =0$, where the system transmission characteristics are obtained from Eq. (3), as shown in Fig. 2.

\begin{figure}[htbp]
 \centering
  \includegraphics[width=8.3cm]{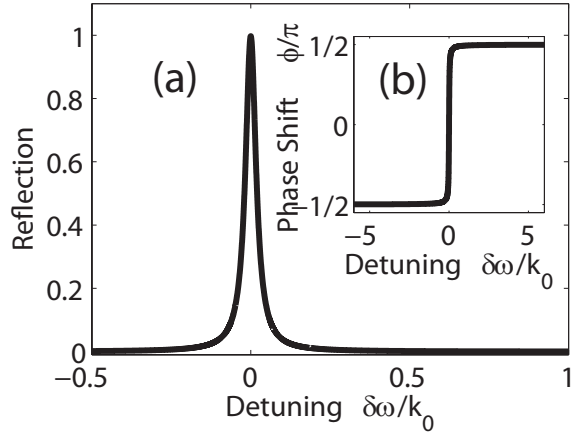}
  \caption{The transmission characteristic of the optical system. (a) The reflection ratio of the input light; (b)Inset: the phase shift of the incident light.}
  \label{fig1}
 \end{figure}
Figure 2(a) shows the relationship between the reflection ratio of the incident light and the dissonance. A sharp reflection peak is obtained at resonance. The full width at half maximum (FWHM) of the peak is around $0.005k_1$, and the reflection ratio is 1, i.e., the system reflects the incident light completely. When there is greater dissonance between the incident light wave and the micro cavity, the reflection ratio decreases to 0, and the system is transparent to the incident light. Figure 2(b) shows that when the incident light and the micro cavity are in the lower dissonance range, the reflected light obtains a $\pi$  phase shift. This is of critical importance to the fabrication of this kind of optical switch, and will be explained in later sections of the paper.

The following sections provide further analysis of the relationship between the transmission characteristics of the entire coupling system and the number of coupled micro cavities and the dissipation coefficients of these micro cavities. $C_{in}^1$ represents the signal light, $C_{in}^2$ represents the control light. The dissipation coefficients of all units are presumed to be equal in this analysis.

First, the impact of the cavity dissipation coefficient on the signal light transmission was analyzed. Three micro cavities were chosen to couple with the waveguide, where $k_2 =25 k_1$, $k_0 = 6k_1$. The relationship between the total transmission and the dissipation coefficient is obtained from Eq. (5), as shown in Fig. 3. Figure 3(a) shows that when the dissipation coefficient of the micro cavity $k_0$ is around $5k_1$ , the transmission ratio of  $C^1_{in}$ is approximately 0.68 for a zero   $C^2_{in}$ input, while the transmission ratio of  $C^1_{in}$  decreases to 0 when the $C^2_{in}$ input is available.

Second, assuming that $C^1_{in}$  should be resonant with the micro cavity, $k_1$ is still used to normalize the other dissipation coefficients, and $k_2 =25k_1$, with $k_0 = 5k_1$. The relationship between the total system transmission and the number of coupled micro cavities is obtained from Eq. (5), as shown in Fig. 3. Figure 3(b) shows that when three micro cavities are coupled to the waveguide, the transmission ratio of $C^1_{in}$  is 0.72 without any $C^2_{in}$ input; again, the $C^1_{in}$  transmission ratio decreases to 0 if an input $C^2_{in}$  is available. If the system is used as an optical switch, the control light $C^2_{in}$ can achieve switched control of the signal light $C^1_{in}$ .
\begin{figure}[htbp]
  \centering
 \includegraphics[width=10.6cm]{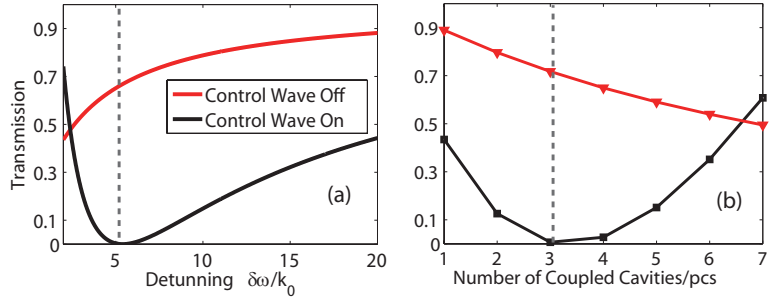}
	\caption{The transmission against the deturnning and the number of coupled cavities. }
 \label{2}
\end{figure}

\section{All-optical Switch and Its Time Series Analysis}
The structure of the model system studied in this paper shows that the $N$ micro cavities are coupled to the optical waveguide L (Fig. 1(a)). Figure 1(b) shows a single-unit structure which would be one part of the system. Consider each micro cavity to be a single cavity, and their natural frequencies are represented by $ \omega_0 $. The other parameters of the model are explained as follows: $k_0$ refers to the dissipation coefficient of the ith cavity coupled with the external environment, and its relationship to the quality factor of the micro cavity, $Q_0$ is $k_0=\frac{2 \omega_0}{Q_0}$. $k_1$and $k_2$ are the dissipation coefficients in the left and right directions, respectively, resulting from the ith cavity coupled to the waveguide. Each micro cavity is coupled with a waveguide through $k_1$ and $k_2$, rather than by direct coupling between each micro cavity. $C^1_{in}$ and $C^2_{in}$ refer to the light wave flow operators of the ith unit inputs from the two sides of the waveguide, and $C^1_{out}$ and $C^2_{out}$ refer to the light wave flow operators of the ith unit outputs from the two sides of the waveguide. 

To control the signal light effectively, three indirectly coupled micro cavities with the same dissipation coefficients and resonance frequencies are coupled with the optical waveguide. The dissipation coefficients of each micro cavity, $k_0$ and $k_2$, are given by $k_0=6k_1$ and $k_2=20k_1$, respectively. The switching characteristics are shown in Fig. 3. The incident signal light$C^1_{in}$ is the output from $C2$ with a transmission ratio of 0.72 when there is no control light input at the $C2$ side (Fig. 3(a)). The incident signal light $C^1_{in}$ is blocked completely when a control light is input at the $C2$ side (Fig. 3(b)) and its transmission ratio equals 0. The control signal light $C^2_{in}$  is thus used to achieve the switching function for the objective signal light $C^1_{in}$. Based on the structural parameters given above, the dissipation of the micro cavity to the two side waveguides is asymmetric ($k_2 =20k_1$). The current system can therefore be regarded as an optical device based on asymmetrical coupling. Analysis of Fig. 2 has shown that the signal light $C^1_{in}$ input from $C1$ penetrates the system with a transmission ratio of 0.72, the control light $C^2_{in}$ incident from $C2$ with a reflection ratio of 0.72 is reflected by the micro cavity, and a $\pi$ phase difference is obtained. The two beams with their phase difference of $\pi$ are superimposed and destructively interfere, and thus no light signal output can be achieved on the $C2$ side. The time series relationship of the optical switch is shown in Fig. 4. 

The results of the analysis above show that when $C^1_{in}$ is used as the objective light signal and $C^2_{in}$ is used as the control light signal, this kind of waveguide system coupling with the micro cavity array provides an all-optical switch model. To control the signal light effectively, three indirectly coupled micro cavities with the same dissipation coefficients and resonance frequencies are coupled with the light waveguide. The dissipation coefficients of each micro cavity, $k_0$ and $k_2$, are given by $k_0=6k_1$ and $k_2=20k_1$, respectively. The switching characteristics are shown in Fig. 3. The incident signal light $C^1_{in}$ is output from $C2$ with a transmission ratio of 0.72 when there is no control light input at the $C2$ side (Fig. 3(a)). The incident signal light $C^1_{in}$ is blocked completely when the control light is input at the $C2$ side (Fig. 3(a)), i.e. the transmission is equal to 0. Thus, the control signal light is used to achieve the switching function for the objective signal light $C^1_{in}$. Under the structural parameters proposed above, the dissipation of the micro cavity to the two side waveguides is asymmetric $(k_2 =20 k_1 )$. The current system can thus be regarded as an optical device based on asymmetric coupling. Analysis of Fig. 2 and Fig. 3 indicates that the signal light input from C1 penetrates the system with a transmission ratio of 0.72 (Fig. 3), the control light $C^2_{in}$ incident from $C2$ with a reflection ratio of 0.72 is reflected by the micro cavity, and the $\pi$ phase difference is obtained simultaneously (Fig. 2). These two light beams with their $\pi$ phase difference are superimposed and destructively interfere, and thus, no light signal output can be achieved on the $C2$ side. The time series relationship of the optical switch is shown in Fig. 4.
\begin{figure}[htbp]
   \centering
   \includegraphics[width=8.3cm]{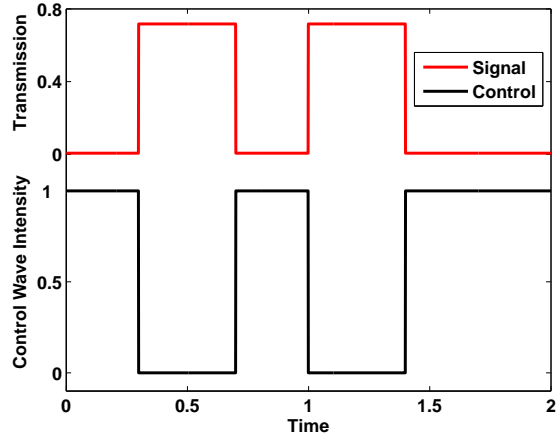}
       \caption{Time series relationship of the optical switch. The red line refers to the signal light, and the black line refers to the control light.}
   \label{fig7}
 \end{figure}

\section{Conclusion}
The analysis of the theoretical results found that highly efficient all-optical switching can be achieved when three micro cavities are used to couple to the optical waveguide and the appropriate dissipation coefficient is chosen. The theoretical model discussed in the current paper can be achieved by introducing point and line defects in photonic crystals, and coupling to the micro disk \cite{barclay2006integration} or micro sphere with an optical fiber \cite{PhysRevLett.92.117902}.

The work in this paper was supported by the National Natural Science Foundation of China (Grant Nos. 11047024 and 11064009), and the Key Project of the Chinese Ministry of Education (Grant No. 211194). The authors would like to thank these two organizations for all their help.

\end{document}